\documentclass{article}
\usepackage{amssymb}

\input{tcilatex}
\begin{document}

\title{Sudden Brans-Dicke Singularities}
\author{John D. Barrow \\
DAMTP, Centre for Mathematical Sciences\\
University of Cambridge, Wilberforce Rd.,\\
Cambrdige CB3 0WA, United Kingdom}
\maketitle

\begin{abstract}
\end{abstract}

\section{\protect\bigskip Introduction}

Following the realisation that finite-time 'sudden' singularities can arise
in general relativistic cosmologies where the scale factor, its first time
derivative and the fluid density can remain finite whilst its second
derivative and fluid pressure diverge although the strong energy condition
remains unviolated \cite{BGT, JB1}, there has been extensive study of this
possibility and its close relatives. Generalisations were found in ref \cite%
{JB2} and examples appeared in anisotropic cosmologies \cite{JB3} and
higher-order gravity theories with $f(R)$ lagrangians \cite{JB2}. These are
weak singularities in the senses of Tipler \cite{T} and Krolak \cite{K} and
their conformal diagrams have been constructed in ref. \cite{dabconf}.
Geodesics are unscathed by sudden singularities, \cite{las} and the general
behaviour of the Einstein and geodesic equations in their neighbourhood was
found in refs. \cite{cot1, cot2, lip}. This behaviour appears robust in the
presence of quantum particle production \cite{q}. The first examples were
existence proofs that required unmotivated pressure-density relations, but
more recently generalised singularities of this sort have been found by
Barrow and Graham \cite{BGrah} to appear in simple Friedmann universes with
a scalar field having power-law self-interaction potential $V(\phi
)=V_{0}\phi ^{n}$, $0<n<1\ $\ which always develop a finite-time singularity
where the Hubble rate and its first derivative are finite, but its second
derivative diverges. For non-integer $n>1,$ there is a class of models with
even weaker singularities. Infinities first occur at a finite time in the $%
(k+2)^{th}$ time derivative of the Hubble expansion rate, where $k<n<k+1$
and $k$ is a positive integer \cite{BGrah}. These models inflate but
inflation ends in a singular fashion.

In this paper we investigate what happens in Brans-Dicke theory \cite{BD},
which generalises general relativity by admitting the possibility of
spacetime variation in the Newtonian gravitational 'constant', $G$. We show
that sudden singularities can also appear in such theories and we display
the effects that they have on on the time-evolution of $G(t)$. Other
investigations of the effect finite-time singularities on varying constants
have been made \cite{dab} in the context of the BSBM theory for varying fine
structure 'constant' \cite{bsbm}, but not with varying $G$, although the two
could be combined \cite{bsm}.

\section{Brans-Dicke Sudden Singularities}

We assume the spacetime metric to be of isotropic and homogeneous form, with
expansion scale factor $a(t)$, where $t$ is the comoving proper time
coordinate, and $r$ is the comoving radial distance and $k$ takes values $0$
or $\pm 1$ depending on the curvature of the space sections of constant time
and the speed of light is set to unity:

\begin{equation}
ds^{2}=dt^{2}-a^{2}(t)\{\frac{dr^{2}}{1-kr^{2}}+r^{2}(\sin ^{2}\theta +d\phi
^{2})\}  \label{metric}
\end{equation}

In standard notation, the field equations of Brans-Dicke theory with
energy-momentum tensor $T_{ab}$ and Brans-Dicke scalar field $\phi (t)$ and
scalar field coupling constant $\omega $ are: 
\begin{eqnarray}
G_{ab} &=&\frac{8\pi }{\phi }T_{ab}+\frac{\omega }{\phi ^{2}}(\phi _{,a}\phi
_{,b}-\frac{1}{2}g_{ab}\phi ^{,c}\phi _{,c})+\frac{1}{\phi }(\nabla
_{a}\nabla _{b}\phi -g_{ab}\square \phi )  \label{bd1} \\
\square \phi &=&\frac{8\pi }{3+2\omega }T_{a}^{a}  \label{bd2} \\
T_{a;b}^{b} &=&0  \label{bd3}
\end{eqnarray}%
The Einstein-Brans-Dicke field equations for Friedman universe with metric (%
\ref{metric}) containing fluid with pressure $p$ and density $\rho $ are :

\begin{eqnarray}
3\frac{\ddot{a}}{a} &=&-\frac{8\pi }{(3+2\omega )\phi }\left[ (2+\omega
)\rho +3p(1+\omega )\right] -\omega \frac{\dot{\phi}^{2}}{\phi ^{2}}-\frac{%
\ddot{\phi}}{\phi }  \label{1} \\
\ddot{\phi}+3\frac{\dot{a}}{a}\dot{\phi} &=&\frac{8\pi }{(3+2\omega )}(\rho
-3p)  \label{2} \\
\dot{\rho}+3\frac{\dot{a}}{a}(\rho +p) &=&0  \label{3} \\
\frac{\dot{a}^{2}}{a^{2}} &=&\frac{8\pi \rho }{3\phi }-\frac{k}{a^{2}}-\frac{%
\dot{\phi}\dot{a}}{\phi a}+\frac{\omega }{6}\frac{\dot{\phi}^{2}}{\phi ^{2}}
\label{4} \\
G &=&\left( \frac{2\omega +4}{2\omega +3}\right) \phi ^{-1}  \label{5}
\end{eqnarray}

We are interested first in a sudden singularity occurring where $\phi ,\dot{%
\phi},a,\dot{a},\varrho $ are finite, but where $\ddot{\phi},p,\ddot{a},\dot{%
\rho}$ can be infinite. Henceforth, we drop the curvature term ($k/a^{2}$)
and take the flat geometry with $k=0$ since the curvature turns out to play
no essential role in the discussion. In principle, singularities in second
time-derivatives of the Brans-Dicke scalar field, $\ddot{\phi}$, could occur
at a different time to those in $\ddot{a}$ and $p$, but it is easy to show
(as we will see below) that all finite-time singularities of this type have
to occur at the same time\footnote{%
In the general solution with a sudden singularity present we have shown that
the form of the solution for the scale factor, eq. (\ref{12}) is generalised
so that the constants $t_{s}$ and $a_{s}$ become functions of the space
coordinates \cite{cot1}. Therefore, the sudden singularity is no longer
simultaneous everywhere but the time evolution on approach to it is the same
everywhere.}, which we label $t_{s}$. At such a finite-time sudden
singularity we see that all terms in (\ref{4}) are finite as $t\rightarrow
t_{s}$ from \ below and the dominant divergent terms in the remaining
equations give the asymptotic system:

\begin{eqnarray}
3\frac{\ddot{a}}{a} &\rightarrow &-\frac{24\pi p(1+\omega )}{(3+2\omega
)\phi }-\frac{\ddot{\phi}}{\phi }  \label{6} \\
\ddot{\phi} &\rightarrow &\frac{-24\pi p}{(3+2\omega )}  \label{7} \\
\dot{\rho} &\rightarrow &-3\frac{\dot{a}}{a}p  \label{8}
\end{eqnarray}%
This system of three equations ensures the sudden singularities in $a,p$ and 
$\phi $ occur at the same time. From (\ref{6}) and (\ref{7}), eliminating $p$%
, we have a consistency relation:

\begin{equation}
3\frac{\ddot{a}}{a}\rightarrow -\frac{24\pi p(1+\omega )}{(3+2\omega )\phi }-%
\frac{\ddot{\phi}}{\phi }=\frac{\ddot{\phi}}{\ 2\phi }(3+2\omega )-\frac{3%
\ddot{\phi}}{2\phi }  \label{9}
\end{equation}%
and so, as $t\rightarrow t_{s}$,

\begin{equation}
\frac{\ddot{a}}{a}\rightarrow \frac{\omega \ddot{\phi}}{3\phi }.  \label{con}
\end{equation}%
This requires the singularities in second derivatives of $a(t)$ and $\phi (t)
$ to be simultaneous. We pick the following forms for the $a(t)$ and $\phi
(t)$ evolution :

\begin{equation}
\phi =\phi _{s}\left( \frac{t}{t_{s}}\right) ^{r}-C\left( 1-\frac{t}{t_{s}}%
\right) ^{n}  \label{11}
\end{equation}

with $0<r<1<n<2$, and

\begin{equation}
a(t)=\left( \frac{t}{t_{s}}\right) ^{q}(a_{s}-1)+1-\left( 1-\frac{t}{t_{s}}%
\right) ^{\lambda }  \label{12}
\end{equation}

with $0<q<1<\lambda <2.$ Hence, as $t\rightarrow t_{s}$ we have

\begin{eqnarray}
a &\rightarrow &a_{s}+q(1-a_{s})(1-\frac{t}{t_{s}})\rightarrow a_{s},
\label{asy1} \\
\phi &\rightarrow &\phi _{s}(1-r[1-\frac{t}{t_{s}}])\rightarrow \phi _{s},
\label{asy2} \\
\frac{\ddot{\phi}}{\phi } &\rightarrow &-\frac{Cn(n-1)}{t_{s}^{2}}(1-\frac{t%
}{t_{s}})^{n-2}\rightarrow \infty ,  \label{asy3} \\
\frac{\ddot{a}}{a} &\rightarrow &-\frac{\lambda (\lambda -1)}{a_{s}t_{s}^{2}}%
(1-\frac{t}{t_{s}})^{\lambda -2}\rightarrow \infty ,  \label{asy4}
\end{eqnarray}

\bigskip The consistency condition (\ref{con}) is satisfied if we take: 
\begin{eqnarray}
n &=&\lambda ,  \label{a1} \\
C &=&\frac{3}{\omega a_{s}}.  \label{a2}
\end{eqnarray}

Therefore, the final form of the solution with the required simultaneous
sudden singularity in $a(t)$ and $\phi (t)$ is:

\begin{equation}
\phi =\phi _{s}\left( \frac{t}{t_{s}}\right) ^{r}-\frac{3}{\omega a_{s}}%
\left( 1-\frac{t}{t_{s}}\right) ^{n},  \label{phi}
\end{equation}

\begin{equation}
a(t)=\left( \frac{t}{t_{s}}\right) ^{q}(a_{s}-1)+1-\left( 1-\frac{t}{t_{s}}%
\right) ^{n},  \label{a}
\end{equation}%
with $1<n<2$. We note that $\frac{\ddot{a}}{a}<0$ and so the strong energy
condition is still satisfied. We note also that if we wish to shift the
sudden singularities up to appear in derivatives of $a$ and $\phi $ that are
higher than second then this can be arranged choosing the range for $n$
suitably, with $r-1<n<r$ in order to have a generalised sudden singularity 
\cite{JB2} that creates an infinity in the $r^{th}$ time derivatives of $a(t)
$ and $\phi (t)$. Other varieties of sudden singularity (see ref(\cite{od})
for a classification of types of finite-time singularity involving
infinities in different combinations of cosmological variables in theories
other than Brans-Dicke) can also be engineered by suitable choice of these
parameters and their ranges.

If at early times, $t\rightarrow 0$, the solution behaves like the special
Brans-Dicke exact solutions with 'Machian' initial condition $\phi (0)=0,$
then

\begin{eqnarray}
a &\propto &\left( \frac{t}{t_{s}}\right) ^{q},  \label{mach1} \\
\phi  &\propto &\left( \frac{t}{t_{s}}\right) ^{r}.  \label{mach2}
\end{eqnarray}%
The relations between $q$ and $r$ that exist for the exact power-law
Brans-Dicke solutions of (\ref{4}) in the presence of a perfect-fluid source
with equation of state \cite{nar}

\begin{equation}
p=(\gamma -1)\rho  \label{eqstate}
\end{equation}%
are

\begin{equation}
3\gamma q+r=2  \label{nar}
\end{equation}

This is the requirement that $\rho /\phi \propto t^{-2}$, and all terms in
the Friedmann-like eq. (\ref{4}) fall as $t^{-2}$ since $\rho \propto
a^{-3\gamma }$ from eq. (\ref{3}). For example, in the case of radiation ($%
\gamma =4/3$) we have $\phi $ constant and $a\varpropto t^{1/2},\rho
\varpropto t^{-2},$ so no variation of $G$ in this solution; in the case of
dust ($\gamma =1$), we have $\phi \varpropto t^{r}$ and $a\varpropto
t^{(2-r)/3},\rho \varpropto t^{r-2}$, where $r$ is arbitrary. The choice of
variation in $G$ can be made slow enough by choice of $r$ to ensure
agreement with big bang nucleosynthesis if required \cite{barrG, clif}.

However, these 'Machian' solutions are not the general solutions of eqs. (%
\ref{1})-(\ref{4}). If we take the general solution of eqs. (\ref{1})-(\ref%
{4}), \cite{rus, JB4}, then $\phi (0)\neq 0$ and the ('non-Machian')
solution is dominated by the scalar field dynamics, rather than by the
matter term, $\rho /\phi $, as $t\rightarrow 0$. In that case we have $%
a\propto t^{(1-\beta )/(3-\beta )\text{ }}$and $\phi \propto t^{2\beta /3}$
with $\beta \equiv \left( \frac{3}{2\omega +3}\right) ^{1/2}>0$ and approach
to the vacuum solution of O'Hanlon and Tupper \cite{Tup}$.$ For large $%
\omega $, on approach of the theory to general relativity, we have $\beta
\rightarrow 0,$and hence, 
\begin{equation}
a\propto t^{1/3};\text{ \ \ \ \ \ }\phi \propto t^{2\beta /3},  \label{vac}
\end{equation}%
and so we have $q=1/3$ and $r=2\beta /3$ for the possible early time
behaviour in general if the vacuum stresses dominate, as we would expect.

As $t\rightarrow t_{s}$, we have the asymptotic forms \emph{\ }

\begin{eqnarray}
a(t) &\rightarrow &a_{s}+q(1-a_{s})(1-\frac{t}{t_{s}}),  \label{f1} \\
\phi (t) &\rightarrow &\phi _{s}\left[ 1-r(1-\frac{t}{t_{s}})\right] ,
\label{f2} \\
G(t) &\rightarrow &\left( \frac{4+2\omega }{3+2\omega }\right) \phi
^{-1}\rightarrow \frac{G_{s}}{1-r(1-\frac{t}{t_{s}})}\rightarrow G_{s}\left[
1+r(1-\frac{t}{t_{s}})\right] ,  \label{f3} \\
\frac{\dot{\phi}}{\phi }\  &=&-\frac{\dot{G}}{G}\rightarrow \frac{r\left[
1-(r-1)(1-\frac{t}{t_{s}})\right] }{t_{s}\left[ 1-r(1-\frac{t}{t_{s}})\right]
}\rightarrow \frac{r}{t_{s}}.  \label{f4}
\end{eqnarray}%
So could an observational bound on $\dot{G}/G$ today can tell us how close
we could be to $t_{s}$? Present-day observations bound $r$ as $\dot{G}/G\sim
r/t_{0}<10^{-12}yr^{-1}$ when $t_{0}<<t_{s}$.The usual power-law fall-off in 
$G$ tails off to a constant value, $G_{s}$ which is smaller than the present
value by a factor $t_{0}/t_{s}$.

\section{More General Situations}

It is straightforward to see the consequences of generalising from
Brans-Dicke theory, where the coupling parameter, $\omega $, is constant, to
a scalar-tensor gravity theory where $\omega =\omega (\phi ),$ as described
in refs. \cite{JB4,BPar}. The essential field equations \ref{(1)-(2)} are
generalised in this case to become \cite{nord}:

\begin{eqnarray}
3\frac{\ddot{a}}{a} &=&-\frac{8\pi }{(3+2\omega )\phi }\left[ (2+\omega
)\rho +3p(1+\omega )\right] -\omega \frac{\dot{\phi}^{2}}{\phi ^{2}}-\frac{%
\ddot{\phi}}{\phi }-\frac{\omega ^{\prime }\dot{\phi}^{2}}{2(3+2\omega )\phi 
}  \label{1W} \\
\ddot{\phi}+3\frac{\dot{a}}{a}\dot{\phi} &=&\frac{8\pi }{(3+2\omega )}(\rho
-3p)-\frac{\dot{\phi}^{2}\omega ^{\prime }(\phi )}{(3+2\omega )}  \label{2W}
\end{eqnarray}

We can see that the appearance of the new non-zero $\omega ^{\prime }(\phi )$
terms does not affect the finite-time singularities created by the
divergences of the $\ddot{\phi}$ and $\ddot{a}$ terms because the $\phi $
and $\dot{\phi}$ terms that multiply them tend to constants as $t\rightarrow
t_{s}$ at the sudden singularity. Hence, we expect the behaviour at sudden
singularities in general scalar-tensor theories to be as described above for
Brans-Dicke. Infinities can occur in $\omega (\phi )$ at some finite time
(even the present day) but are usually harmless. The $\omega (\phi
)\rightarrow \infty $ limit is part of the general relativity limit. The
other requirement in this limit is that\footnote{%
The requirement is $\left\vert \frac{\omega ^{\prime }}{(3+2\omega
)^{2}(4+2\omega )}\right\vert \rightarrow 0$ when $\omega \rightarrow \infty 
$. In that limit this reduces to $\omega ^{\prime }/\omega ^{3}\rightarrow 0.
$} $\omega ^{\prime }/\omega ^{3}\rightarrow 0$ \ as $\omega \rightarrow
\infty $. So, for example, if $\omega (\phi )\varpropto (\phi -\phi
_{0})^{n},\cite{BPar}$, then we require $n<0$ for $\omega \rightarrow \infty 
$ as $\phi \rightarrow \phi _{0}$ and $n\leqslant -1/2$ to ensure $\omega
^{\prime }/\omega ^{3}\rightarrow 0.$ We have not included the potential
term \cite{bm} in the general scalar-tensor theory and expect new features
will enter with its presence when its form is suitably chosen. The effects
will mirror those of power-law scalar field potentials in general relativity
found in ref \cite{BGrah} and lead to infinities in higher than second
powers of the scale factor.

When a self-interaction potential, $V(\phi )$ is also present in these
theories, it adds terms of the form

\begin{equation}
\frac{1}{2\omega +3}\left( \phi V^{\prime }(\phi )-2V(\phi )\right) .
\label{V}
\end{equation}

to the right-hand side of eq. (\ref{2W}). Therefore, we expect the new
higher-order singularities found by Barrow and Graham \cite{BGrah} in
general relativity with power-law scalar field potentials to occur also for
scalar-tensor cosmologies with $V(\phi )\varpropto \phi ^{n},$ $n\neq 2.$

\section{Discussion}

We have extended the study of finite-time singularities of 'sudden' type
from general relativity and associated $f(R)$ gravity theories to
scalar-tensor theories which incorporate varying $G$. We have constructed
the form of these singularities in Brans-Dicke gravity theory and argue that
more general scalar-tensor theories introduce no new features in the absence
of potentials. We have shown elsewhere \cite{cot1} that the sudden
singularity forms for the scale factor as $t\rightarrow t_{s}$ are generic
to general relativistic cosmology in the sense that if the form (\ref{12})
is generalised, so that the constants become space functions of space and
the terms form the leading ones in a series expansion around the
singularity, then the resulting solution has the required 9-function spatial
arbitrariness required of a general solution of the Einstein equations
without an imposed equation of state. We expect the same arguments to hold
for the asymptotic form (\ref{12}) in Brans-Dicke theory and our
constructions will form part of the general solution of the field equations
in scalar-tensor theories also.  In \cite{cot2}, we considered the general
behaviour of the geodesics on approach to a general sudden singularity of
the form (\ref{12}). The geodesic behaviour is general but it remains an
unresolved question as to whether the fact that geodesics are blind to these
singularities, as first shown in ref. \cite{las}, indicates something
unphysical about them. However, we recognise the existence of finite-time
singularities in other areas of physics in flat spacetime and the sudden
singularities studied here and elsewehere might just be regarded as the
general relativistic or scalar tensor theoretic editions of these in curved
spacetime.  

\textbf{Acknowledgment}: We acknowledge support from the STFC of the UK.


\begin{thebibliography}{99}
\bibitem{BGT} Barrow J D, Galloway G and Tipler F J 1986 \textit{Mon. Not.
R. Astron. Soc. }\textbf{223} 835

\bibitem{JB1} Barrow J D 2004 \textit{Class. Quantum Gravity }\textbf{21} L79

\bibitem{JB2} Barrow J D 2004 \textit{Class. Quantum Gravity }\textbf{21}
5619

\bibitem{JB3} Barrow J D and Tsagas C G 2005 \textit{Class. Quantum Gravity }%
\textbf{22} 1563

\bibitem{T} Tipler F J 1977 \textit{Phys. Lett. A} \textbf{64} 8

\bibitem{K} Krolak A 1986 \textit{Class. Quantum Gravity }\textbf{3} 267

\bibitem{dabconf} D\c{a}browski M P and Marosek K 2018 arXiv:1806.00601

\bibitem{las} Fernandez-Jambrina L and Lazkoz R 2004 \textit{Phys. Rev. D} 
\textbf{70} 121503(R)

\bibitem{cot1} Barrow J D, Cotsakis S T and Tsokaros A 2010  \textit{Class.
Quantum Gravity} \textbf{27}165017

\bibitem{cot2} Barrow J D, and Cotsakis S T 2013\textit{\ Phys. Rev. D} 
\textbf{88} 067301

\bibitem{lip} Barrow J D and Lip S Z W 2009 \textit{Phys. Rev. D} \textbf{80}
043518

\bibitem{q} Barrow J D, Batista A B, Fabris J C and Houndjo S 2008 \textit{%
Phys. Rev. D }\textbf{78} 123508

\bibitem{BGrah} Barrow J D, and Graham A A H 2015 \textit{Phys. Rev. D} 
\textbf{91} 083513

\bibitem{dab} D\c{a}browski M P, Denkiewicz T, Martins C J A P and Vielzeuf
P E 2014 \textit{Phys. Rev.} \textit{D }\textbf{89} 123512

\bibitem{bsbm} Sandvik H, Barrow J D and Magueijo J 2002 \textit{Phys. Rev.
Lett.} \textbf{88} 031302

\bibitem{bsm} Barrow J D, Magueijo and J Sandvik H 2002 \textit{Phys. Lett. B%
} \textbf{541} 201

\bibitem{BD} Brans C H and Dicke R H 1961\textit{\ Phys. Rev.} \textbf{124}
925

\bibitem{nar} Nariai H 1969 \textit{Prog Theo. Phys.} \textbf{42} 544

\bibitem{bm} Barrow J D and Maeda K-I 1990 \textit{Nucl. Phys. B} \textbf{341%
} 294

\bibitem{barrG} Barrow J D 1978. \textit{Mon. Not. Roy. astr. Soc.} \textbf{%
184} 677.

\bibitem{clif} Clifton T, Barrow J D,  and Scherrer R J 2005 Phys.Rev. D 
\textbf{71} 123526 

\bibitem{rus} Gurevich L E, Finkelstein A M and Ruban V A 1973 \textit{%
Astrophys. Space Sci.} \textbf{22} 231

\bibitem{Tup} O'Hanlon J and Tupper B O J 1970 \textit{Nuovo Cimento} 
\textbf{137} 305

\bibitem{JB4} Barrow J D 1992 \textit{Phys. Rev. D} \textbf{47} 5329

\bibitem{od} Nojiri S and Odintsov S D 2008 \textit{Phys.Rev.D} \textbf{78}
046006

\bibitem{BPar} Barrow J D and Parsons P 1997 \textit{Phys. Rev. D }\textbf{55%
} 1906

\bibitem{nord} Nordtvedt K 1970 Ap J \textbf{161} 1059
\end{thebibliography}
\end{document}